# Towards a Formalization of the Unified Modeling Language[*]

Ruth Breu, Ursula Hinkel, Christoph Hofmann, Cornel Klein,
Barbara Paech, Bernhard Rumpe, Veronika Thurner

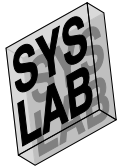

Institut für Informatik
Technische Universität München
D-80290 München
http://www4.informatik.tu-muenchen.de/

**Abstract.** The Unified Modeling Language UML is a language for specifying, visualizing and documenting object-oriented systems. UML combines the concepts of OOA/OOD, OMT and OOSE and is intended as a standard in the domain of object-oriented analysis and design. Due to the missing formal, mathematical foundation of UML the syntax and the semantics of a number of UML constructs are not precisely defined. This paper outlines a proposal for the formal foundation of UML that is based on a mathematical system model.

## 1 Introduction

The Unified Modeling Language [2] is a set of description techniques suited for specifying, visualizing and documenting object-oriented systems. The language has been developed by G. Booch, J. Rumbaugh and I. Jacobson since October 1994 and combines the concepts of OOA/OOD [1], OMT [24], and OOSE [17], as well as a number of ideas from other methods and description techniques like Harel's statecharts [14].

In January 1997 UML has been submitted to OMG as a proposal for a standard notation of object-oriented analysis and design techniques [2]. Currently, UML focuses only on notation. Method and process issues are outlined, but not dealt with in detail. However, it is stated that the process is to be use-case driven, architecture centric, iterative and incremental (Summary of [2], p. 7). In our work, we refer to the most recent UML version 1.0.

Like other software engineering methods UML provides a set of "intuitive" graphical and textual description techniques that are supposed to be easily understandable for both system developers and expert users working in the application domain. However, often the exact meaning of such description techniques is not clearly defined. As a consequence, the usage of those techniques and, correspondingly, the interpretation of models developed may differ considerably. Furthermore, without exact semantics, checks for completeness and consistency cannot be precisely defined, let alone supported by a tool. Quite often, the models emerging during system development have severe shortcomings, which inevitably lead to erroneous software systems. Therefore, the high effort spent on modeling not always yields software systems of high quality.

---

[*] This paper partly originates from a cooperation of the DFG project Bellevue and the SysLab project, which is supported by the DFG under the Leibniz program, by Siemens-Nixdorf and Siemens Corporate Research.

In order to ensure the correct usage of description techniques in modeling, and to enable tool supported consistency checks, the definition of a precise semantics of the notations involved is crucial. The semantics defines the exact meaning of description techniques in an unambiguous way. Furthermore, the formal framework serves as a basis for defining the interconnections between different notational concepts and different stages of design. Last but not least a semantic foundation checks the soundness of the description technique and thus may lead to an improvement of the description technique itself.

Having recognized the importance of a formal foundation, the UML developers already have made first attempts at a formal semantics definition. In the language documentation a metamodel for UML concepts is presented. The metamodel itself is given in UML notation by a class diagram and annotations in prose. This approach to a formal semantics of UML brings about several difficulties.

First, the semantics of class diagrams is not precisely defined itself. For example, the usage of aggregation is a frequently discussed topic. Consequently, class diagrams provide a very weak basis for defining a formal semantics.

Second, the use of class diagrams limits the semantics definition to a description of static relationships between UML concepts. As a documentation of the structure of diagrams, the UML metamodel contains valuable information for tool developers who have to handle storage and retrieval of diagrams. However, there exists no interpretation that models the dynamic aspects of system behavior in an appropriate way. Thus, the metamodel is not sufficient as a formal semantics definition of UML concepts. As far as we know, also the novel approaches to a semantics definition pursued by the UML developers [20] do not overcome this deficiency.

Our approach to the formal foundation of UML is based on the well-studied and established mathematical theory of streams and stream processing functions [5]. Streams have proved to be an adequate setting for the formalization of the semantics of concurrent systems. In order to model the static and dynamic properties of an object-oriented system in a structured way, we augment the mathematical framework by the notion of *system models*. A system model characterizes an abstract view of the systems under development. A system model both describes the static structure of objects and their behavior over time. The idea of a system model is advantageous for several reasons.

First, a system model provides an integrated view of a system. This is particularly important as the UML description techniques allow us to define only partial views of a system. The semantic mapping of partial syntactical system views to an overall mathematical system view has the advantage that relationships between different description techniques can be studied in a homogeneous setting.

Second, the concept of system models establishes an auxiliary layer on top of the basic mathematical theory. In this semantic layer object-oriented notions like objects, object identities and object states have a direct correspondence to mathematical concepts in the system model. Thus, the use of a system model helps us to increase the readability and understandability of the semantics definition considerably.

For the semantics definition we employ our experience gained during the SYSLAB project. In SYSLAB a formally founded design method has been developed covering description techniques similar to those of UML [18, 13, 12, 15, 21, 29].

The intention of this paper is to outline the basic ideas and the overall structure of the formal foundation of UML. Through the semantic definition of UML concepts, we detected a number of language features which are not yet fully clear. We discuss some of these aspects in the respective sections.

The paper is organized as follows: In Section 2 we give a short overview of the basic modeling concepts of UML. In Section 3 we present a proposal for the formal foundation of UML. The subsections of Section 3 focus on the overall mathematical system view and on the different description techniques UML offers. Section 4 contains a summary and our conclusions.

## 2 A Short Overview of UML

In the following we give a short sketch of the basic UML description techniques:

- class and object diagrams,
- use case diagrams,
- sequence diagrams,
- collaboration diagrams,
- state diagrams and
- activity diagrams.

Note that we concentrate on those models and description techniques that are relevant for describing the structure and behavior of systems. Therefore, we omit the implementation diagrams (component and deployment diagrams), which are helpful for modeling the physical structure of a system only. Furthermore, we focus on basic concepts, but omit some more advanced modeling features which are beyond the scope of this work. For a detailed description of UML we refer to [2].

**Class and object diagrams:** A class diagram describes the static structure of a system, consisting of a number of classes and their relationships. A class is a description of a set of objects and contains attributes and operations. An object diagram is a graph of instances. A static object diagram shows the detailed state of a system at a certain point in time, whereas a dynamic object diagram, also called collaboration diagram, models the state of a system over some period of time.

Structural relationships between objects of different classes are represented by associations (the instances of associations are called links). The definition of associations may be enhanced by attributes, association classes, role names and cardinality (multiplicity). Generalization represents the relationship between superclasses and subclasses, i.e. between a more general class and a more specific class. Thus, the specific class is fully consistent with the superclass and adds additional information. Aggregation, which is a concept of OMT, is a special form of binary association representing the whole-part relationship. Composition is a form of aggregation for n-ary associations, which implies strong ownership and coincident lifetime of a part with the whole. For structuring complex systems, class packages are introduced, which are groupings of class model elements and may be nested.

The **use case diagram** captures Jacobson's use cases. A use case diagram shows a collection of use cases and external actors that interact with the system. A use case describes the interactions and the behavior of a system during an entire transaction that involves several objects and actors. Within a use case model, relationships between use cases can be modeled, i.e. a use case can include other use cases as part of its behavior description. The specification of the external behaviour of a use case may be given by a state diagram. The implementation of a use case can be described by a collaboration diagram.

Since the use case diagram is strongly connected with the development process, we omit it in the current stage of our semantics definition.

**Sequence diagrams**, called interaction diagrams in OOSE, show patterns of interactions (i.e. the sending of messages) among a set of objects in a temporal order. In addition, a sequence diagram may show the lifelines of the objects involved in the interactions.

**Collaboration diagrams** are similar to object diagrams in OOA/OOD and describe the collaboration between objects. Collaboration diagrams depict objects and links between them. Links visualize the message flow between the corresponding objects. Messages may have an argument list and a return value. Message ordering in the overall transaction is described by a modified Dewey decimal numbering, specifying the sequential position of a message within its corresponding thread. A composite object is an instance of a composite class which implies the composition aggregation betweeen the class and its part. Parameterized collaborations represent design patterns that can be used repeatedly in different designs.

**State diagrams**, based on the statecharts by Harel [14], are similar to the state-machine diagrams used in OOA/OOD and OMT. They describe the reaction of an object, in reply to events received, in form of responses and actions.

State diagrams basically consist of states and state transitions. A state represents a condition during the existence of an object in which it waits for an event to be received, performs some action or satifies some condition.

An event is an occurrence that may trigger a state transition. Examples for events are the receipt of an explicit signal, and the call upon an object's method. State transitions describe which events an object can receive in a particular state and which state the object adopts after the reception of the event. The sending of events to other objects is part of the transition.

An additional concept in state diagrams are atomic and non-interruptible actions, which are connected to a transition. An action is executed when the corresponding transition fires. It is also possible to invoke internal "do" actions that are carried out within a state and take time to complete. An internal action is initiated when the state is entered and can be interrupted by an event that triggers a state transition.

Timing conditions on the behavior of an object can be introduced by transition times that are associated with a transition to specify the time at which the transition is to fire. Like in statecharts, nesting of states is specified by introducing concurrent or mutually exclusive disjoint substates.

**Activity diagrams** are a special case of state diagrams that are to be used in situations where most of the events represent the completion of internally-generated actions. Thus, the behavior is dominated by internal processing. In contrast, state diagrams are to be used for situations where mainly asynchronous events occur.

An essential feature of UML is the concept of **stereotypes**. Stereotypes are used for classifying modeling elements, thus allowing the user of UML to extend the semantics of the metamodel and to adapt the predefined notational concepts of UML to specific needs. For the evolution of a design the **refinement** relationship associates two descriptions of the same thing at different levels of abstraction. Refinement includes, among others, the relation between an analysis class and a design class.

## 3  A Proposal for the Formal Foundation of UML

This section represents a proposal for a formal foundation of UML. First, we describe our approach to a formalization and introduce the mathematical system model that is used to give an integrated underlying formal semantics for all description techniques of UML. Then, we describe how the semantics of the description techniques of UML can be formalized with respect to the system model.

### 3.1  Roadmap to Formalization

In the introduction we have motivated, *why* a formalization of UML description techniques is useful. We argued that a precise semantics is important not only for the developer, but also for tool vendors, methodologists (people that create the method) and method experts (people that use the method and know it in detail).

Thus, we get the following requirements for a formalization:

1. A formalization must be complete, but as abstract and understandable as possible.
2. The formalization of a heterogeneous set of description techniques has to be integrated to allow the definition of dependencies between them.

This does not mean that every syntactical statement must have a formal meaning. Annotations or descriptions in prose are always necessary for documentation, although they do not have a formal translation. They may eventually be translated into a formal description or even into code during software development when the system model is further refined.

To manage the complexity of formalization, a layer between syntactic description techniques and pure mathematics is introduced, as depicted in Figure 1. The pure mathematics is only used to define the *system model*. This system model is then used as an integrated underlying semantics for all description techniques.

As a further advantage, the system model explicitly defines notions of software systems in terms of mathematical concepts, e.g. object identifiers and messages. In contrast to the more implicit semantics of many other approaches, this leads to a better understanding of the developed systems.

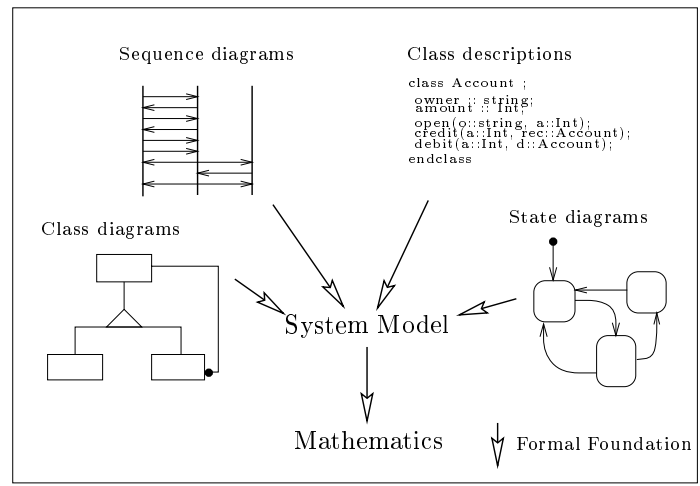

**Fig. 1.** Layered formalization of description techniques

The system model formally defines a notion of a system that obeys the properties defined in Section 3.2. A document of a given description technique is defined by relating its syntactic elements to elements of a system, such as the existing set of classes, or other structural or behavioral entities. The semantics of a document is then given by a subset of the system model. This subset of the system model consists exactly of all systems that are correct implementations of the document.

To use a set of systems and not a single one as the basis of the proposed semantics has several advantages. For example, refinement of documents corresponds to set inclusion. Furthermore, we get the meaning of different documents modeling different aspects of the system by intersection of their respective semantics. But the main reason is that, in contrast to fully executable programming languages, description techniques allow *underspecification* of system properties in many different ways. A proper semantics thus cannot be captured by a single system. For the same reason, it is not possible to give an operational semantics in the sense that a document specifies a single abstract machine that "executes" it.

## 3.2 System Model

The system model described below is a refinement of the SYSLAB system model as presented in [18], [27] and [11]. Each document, for instance an object diagram, is regarded as a constraint on the system model. The system model provides a common basis to define an integrated semantics of all description techniques. On this basis, notions like *consistency* and *refinement* of documents can be precisely defined.

The system model introduced below is especially adapted to the formalization of UML. Thus, relevant aspects of UML like classes, objects, states, messages etc. are explicitly included. A precise formalization of our UML system model is currently under development in [16].

Formally, the *system model* is a set of systems. A *system* is formally described by a tuple of elements that describe various aspects of the system, such as the structure and the behavior of its components as well as their interaction. In the following, we describe the most important elements of a system with identifier *sys*.

The structure of a system is, according to object-orientation, given by a set of objects, each with a unique identifier. Therefore, we regard the enumerable set *ID* of object identifiers as an element of the tuple *sys*.

In the system model objects interact by means of *asynchronous message passing*. Asynchronous exchange of messages between the components of a system means that a message can be sent independently of the actual state of the receiver. Asynchronous system models provide the most abstract system models for systems with message exchange, since deadlock problems as in synchronous systems do not occur. Note that synchronous message passing can be modeled by using two asynchronous messages, a "call" and a "return". To model communication between objects we use the theory of timed communication histories as given in [6]. The notion of explicit time in the system model allows us to deal with real time, as proposed in UML.

We regard our objects as spatially or logically distributed and as interacting in parallel. As described in UML, sequential systems are just a special case, where always exactly one object is "active".

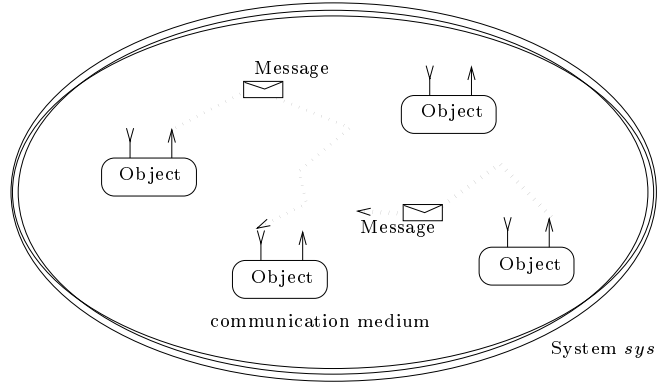

**Fig. 2.** Objects in the UML system model

Interaction between objects occurs through the exchange of messages, as shown in Figure 2. Let *MSG* be an element of *sys*, denoting the set of all possible messages in a system. Each object with identifier $id \in ID$ has a unique set of messages it accepts. Its input interface is defined by

$msg_{id} \subseteq MSG$

The *behavior* of an object is the relationship between the sequences of messages it receives and the sequences of messages it emits as a reaction to incoming messages. We allow our objects to be nondeterministic, such that more than one reaction to an input sequence is possible.

According to [5, 9], the set of timed communication histories over $M$ is denoted by $M^{\overline{\omega}}$. Each communication history contains as information the time unit in which a message occurs, as well as a linear order on the messages it contains. A communication history thus models the observable sequence of incoming or outgoing messages of one object. The behavior of a nondeterministic object $id$ is then given by the mapping of its input stream to the set of possible ouput streams. Thus, the behavior of an object $id$ is given by the relation between its input and output streams

$behavior_{id} \subseteq msg_{id}^{\overline{\omega}} \times MSG^{\overline{\omega}}$

Objects encapsulate data as well as processes. *Encapsulation of data* means that the state of an object is not directly visible to the environment, but can be accessed using explicit communication. *Encapsulation of process* means that the exchange of a message does not (necessarily) imply the exchange of control: each object can be regarded as a separate process. Given the set of possible states *STATE* of objects in a system, the function *states* assigns a subset of possible states to every object:

$states_{id} \subseteq STATE$

Furthermore, a state transition system is associated with each object, modeling the connection between the behavior and the internal state of an object. We use a special kind of automata [13] for this purpose.

Such an automaton of an object *id* consists of a set of input messages $msg_{id}$, a set of output messages *MSG*, a set of states $states_{id}$, and a set of initial states $states_{id}^0 \subseteq states_{id}$. The nondeterministic transition relation $\delta_{id}$ defines the behavior of the automaton. From the state-box behavior, given for the automaton in terms of state transitions, the black-box behavior in terms of the *behavior*-relation can be derived (cf. [13]).

Messages are delivered by a *communication medium*, which is an abstraction of message passing as it is done in real systems by the runtime system of the programming language or by the operating system. The communication medium buffers messages as long as necessary. Each message contains the receiver's identifier, so that the communication medium is essentially composed of a set of message buffers, one for each object. The order of messages between two particular objects is always preserved by the communication medium. The contents of messages are not modified. Messages cannot be duplicated or lost. No new messages are generated by the communication medium. This is formalized in [11].

Objects are grouped into classes. We assume that each system owns a set *CN* of class names. *CN* may, for instance, be derived from UML class diagrams. In object-oriented systems, each object identifier denotes an object that belongs to exactly one class. This is represented by the function

$class : ID \rightarrow CN$.

Classes are structured by an inheritance relation, which we denote by . $\sqsubseteq$ . (read: "subclass of"). The inheritance relation is transitive, antisymmetric and reflexive, as usual. With every class $c \in CN$ a signature $\Sigma_c$ is associated, containing all attributes and methods together with their argument and result types. The signature induces a set of input messages for each object of the class. One impact of inheritance is that signatures are only extended: $c \sqsubseteq d \Rightarrow \Sigma_d \subseteq \Sigma_c$.

Another distinguishing feature of object-orientation is the dynamic creation of objects. Deletion need not be modeled, as we assume that our objects are garbage collected in the usual way. However, we may define a special *finalize()*-method that may be used to clean up objects, as, for instance, in Java. Initially, a finite subset of objects (usually containing one element) exists and is active. We regard objects to be created and to be active after having received a first message. Thus, the creation of a new object essentially consists of a message transmission from the creator to the created object. To allow this, each object is equipped with a sufficiently large (usually infinite) set of object identifiers denoting the set of all object identifiers the object may create:

$$creatables : \mathit{ID} \to \mathcal{P}(\mathit{ID})$$

To prevent multiple creation, these sets of identifiers have to be pairwise disjoint, and objects that are initially active are not creatable at all.

### 3.3 Class and Object Model

Class and object diagrams describe the static structure of a system. The origin of class diagrams are E/R diagrams, which have been successfully applied for years in database design. Although class diagrams are widely accepted in practice, the straightforward adaptation of E/R diagrams to an object-oriented context (through the correspondence *entity = object*) leads to deep semantic problems, since a number of features in E/R diagrams have no exact interpretation in the object-oriented setting. Below, the main concepts and problems of class and object diagrams in UML are summarized, and their mapping to the system model is sketched.

**Classes and Objects** Intuitively, a class $c$ in an UML class diagram describes a set of objects. This is reflected in our system model by three aspects. First, the methods and attributes of class $c$ describe the syntactical interface of all objects belonging to that class. This syntactical interface defines the signature $\Sigma_c$ as given in the system model. Second, the state space of the objects of class $c$ is determined. The state of an object is structurally determined by the attributes of the class and may contain both basic values (like integers or strings) and identifiers of other objects. The set of all states of objects of class $c$ is denoted by $states_{id}$. Third, a subset $ID_c$ of the set $ID$ of all identifiers is defined, although only implicitly by stating $|ID_c| = \emptyset$ for abstract classes, resp. $|ID_c| = \infty$ for others. The set $ID_c$ is the set of all identifiers of objects of class $c$, subclasses not included.

A class diagram describes the object structure of the system to be developed. In this respect, the semantics of the whole class diagram is the set of possible *system states*. A system state consists of the state of all objects that exist at some point in time. Formally, we describe a system state by an indexed family $\{s_{id} : id \in \mathit{ID}, s_{id} \in states_{id}\}$.

**Associations** Associations between classes in UML are supported in various other object-oriented analysis methods and originally come from the notion of relationship types in the entity/relationship approach.

The system view of E/R modeling is based on a global system state and global transactions on the system state. In this setting, relationship types are modeled by entities (set theoretic relations or tables) with the property of bidirectionality and symmetry.

It is obvious that in the object-oriented framework associations have to be interpreted in a different way: both dynamic behavior and states are localized in the objects. There are several alternatives to interpret associations and links in the context of classes and objects. In order to clarify these alternatives, we use the simple example of Figure 3, where we model the distributed structure of a warehouse by two classes *Branch* and *Central Office* connected by an association *coordinates*.

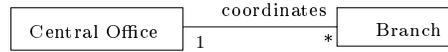

**Fig. 3.** A class diagram modeling a distributed warehouse

- One possibility is to interpret an association as a set of *data links*. In the example this means that a central office object "knows about" branch objects and vice versa. Associations therefore pose additional requirements on the object states. Inherently, associations in this interpretation are not bidirectional relations but correspond to two (semantically independent) unidirectional relations. See for example [25]. The consistency of the two relations is an integrity constraint imposed on linked objects. Another feature related with associations, the specification of their multiplicity, is also an integrity constraint between linked objects and is discussed below.
- A second possibility is to model any association by a separate class, a so-called *association class*. At first sight, this solution seems to be close to the interpretation of relationship types in the E/R approach. However, the paradigm of local object states requires every tuple of linked objects to be connected via an object of the association class. Thus, in this interpretation bidirectionality has to be modeled explicitly and the consistency problem sketched above remains. Thus, this modeling alternative is less abstract in the object-oriented setting than the first alternative and should be limited to the case in which associations are equipped with additional attributes.
- A third solution is to interpret associations as *communication links*. In the example the association *coordinates* then means that a central office object is able to communicate with branch objects and vice versa. Communication links in most cases induce data links, since a prerequisite for communication with other objects is to know about their existence.

In the sense of underspecification, we define the semantics of an association as one of these solutions. The actual choice is left to the developer, e.g. when it becomes clear which objects will send messages along the association. However, in this paper we only talk about the first and simplest solution. In our system model, an association between two classes is modeled within the set of states of the respective objects.

**Object Diagrams** Conceptually, an object icon in an object diagram depicts a single object at a certain point of time (with fixed attribute values). An object diagram thus describes a snapshot of the system and corresponds to a set of system states in our system model. However, the use of an object icon together with class icons usually means that an appropriate object is present in all system states, from beginning to termination. This is formalized by adding an appropriate identifier to the set of initially active objects in the system model.

UML allows some relaxations and extensions of the notations of objects. Among these extensions are the definition of anonymous objects (i.e. objects specified solely by their class without an object identifier), objects without associated attribute values and the stack icon denoting multiple objects.

Anonymous objects stand for "an object" of the given class. Rather than single system states, object diagrams with anonymous objects describe structural properties of system states in a similar way as class diagrams do.

**Aggregation and Composites** UML supports two kinds of aggregation: Shared aggregation and composite aggregation (composition). In a composition, the lifetime of the parts is closely related with the lifetime of the whole. Therefore, "the multiplicity of the aggregate may not exceed 1" ([2], Notation Guide, p. 47), i.e. the parts are not shared among several aggregates. In contrast, shared aggregation puts less constraints on the association, since it allows for sharing, and decouples the lifetimes of the parts from the lifetime of the whole.

This differentiates the current version of UML from Version 0.91, where both concepts have been inconsistently mixed into one. Like constraints, aggregations and compositions are conditions on the system state, and, therefore, can easily be mapped into the system model.

**Constraints** Constraints are conditions on the system state. Constraints can refer to single objects (e.g. for specifying dependencies between attributes) or to several (linked) objects. In UML, constraints are specified as informal text. In order to enable a formal modeling we consider constraints to be predicates over the system states consisting of objects. As already discussed, further types of constraints are induced by other features of class diagrams, e.g. by the multiplicity indicators and by dependencies between associations.

Because there are a lot of different kinds of constraints, a general solution for constraint formalization is not possible. However, the definition of new types of precisely expressible constraints would considerably improve UML. This would allow design decisions regarding static properties of a system to be captured in a more precise and compact way.

**Generalization** Inheritance is the generalization relation between classes. In our system model, inheritance is modeled by $. \sqsubseteq .$ and induces the following three relations:

- Subclasses extend the interface of their superclasses. In our system model this means that the signature of the superclass is a subset of the signature of any of its subclasses.

- A second relation relates the state spaces of super- and subclasses. This structural relation models the property that objects of subclasses have the attributes of their superclasses and participate in associations belonging to their superclasses.
 - A third effect of inheritance concerns the sets of object identifiers. For a given class $c$, the set of associated objects is given by $\{id \in \mathit{ID} \,|\, \mathit{class}(id) \sqsubseteq c\}$. The inheritance relation induces a subset relation between the sets of object identifiers associated with the subclass and the superclass. This subset relation models *(subtype) polymorphism*, i.e. the property that each object of a subtype is also an object of the supertype.

The above relations describe the static properties of super- and subclasses. In the UML documentation nothing is stated about the dynamic properties of inheritance, i.e. how the behaviors of super- and subclasses are related. In fact, inheritance of dynamic behavior is an issue that has been neglected in object-oriented analysis methods so far.

Behavioral inheritance is a well-studied notion at the level of formal specifications (subclasses inherit the abstract properties of their superclasses, see for example [19], [22]) and at the level of programming languages (subclasses may inherit the code of methods of their superclasses). In contrast, only first attempts have been made to relate state diagrams of superclasses and state diagrams of subclasses. One approach to this problem has been presented in [25] and [26].

**Class Packages** Class packages group parts of a class diagram. They define a syntactical name space and, therefore, need no semantic counterpart in the system model.

Class packages may contain classes of other packages that are assumed to be imported. The dependency between class packages can be interpreted as the visualization of such an import of classes. Aggregation of class packages can be seen as the alternative presentation of hierarchically nested packages.

### 3.4 Sequence and Collaboration Diagrams

In contrast to state diagrams, which describe local behavior of objects, sequence diagrams describe global behavior, i.e. interaction sequences between objects. However, the methodological use of sequence diagrams has to be precisely investigated, because sequence diagrams do not provide a complete specification of behavior, but only describe exemplary scenarios. Since collaboration diagrams and sequence diagrams express similar information, but show it in different ways, all propositions made about sequence diagrams in this section apply to collaboration diagrams as well (see [2], Notation Guide, p. 66).

**Exemplary Behavior** The goal of sequence diagrams is to model typical interaction sequences between a set of objects. In Figure 4 a sequence diagram, similar to the one in the UML Notation Guide, is given. The sequence diagram depicts a typical scenario of interaction between the three objects named Caller, Exchange and Receiver.

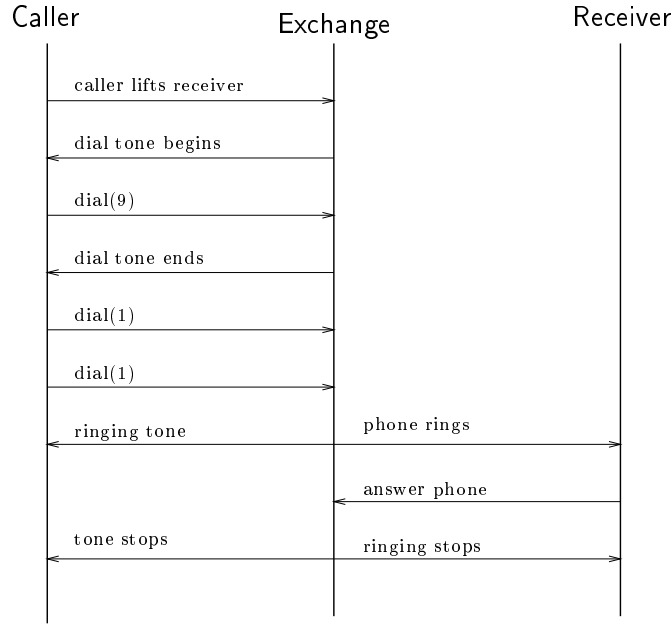

**Fig. 4.** A sequence diagram modeling a phone call

While the concentration on standard cases leads to an easy-to-use notation that is understandable by both software engineers and application experts, it has to be stressed that a sequence diagram does not describe a necessary, but only a possible (or exemplary) interaction sequence between the involved objects. This leads to a semantic problem if sequence diagrams should be considered as a *specification technique*.

In particular, a sequence diagram does not specify in which states the objects have to be in order for the described interaction sequence to occur. For instance, in the above example the phone would not ring if the receiver was busy. Moreover, even if these states had been specified (for instance by giving an interaction sequence leading to the state), the sequence diagram would still leave open whether the described interaction sequence is the only possible one to occur or whether there are other possible interaction sequences. Therefore, from a strictly formal point of view, a sequence diagram not really makes a proposition about the executions of a system.

Note that this is a principal problem that stems from the fact that the objective of sequence diagrams is to describe exemplary behavior. This problem

can be relaxed by using additional language constructs such as repetition and
choice, thus providing a means for the description of complete sets of alternative
sequence diagrams.

We are currently developing a method for a seamless transition from exemplary behavior descriptions that can be expressed, for instance, using sequence
diagrams, to complete specifications using state diagrams.

**Formalization** We formalize sequence diagrams by adopting a state box view.
For each vertical line in a sequence diagram that corresponds to an object an
abstract state automaton is defined along the lines of [12]. State automata consist
of a set of states, an initial state, and a set of transitions. In our case, a transition
is either labeled by an input event or by an output event. State automata can
easily be translated into state transition systems of the system model [12], but
this is not exploited here.

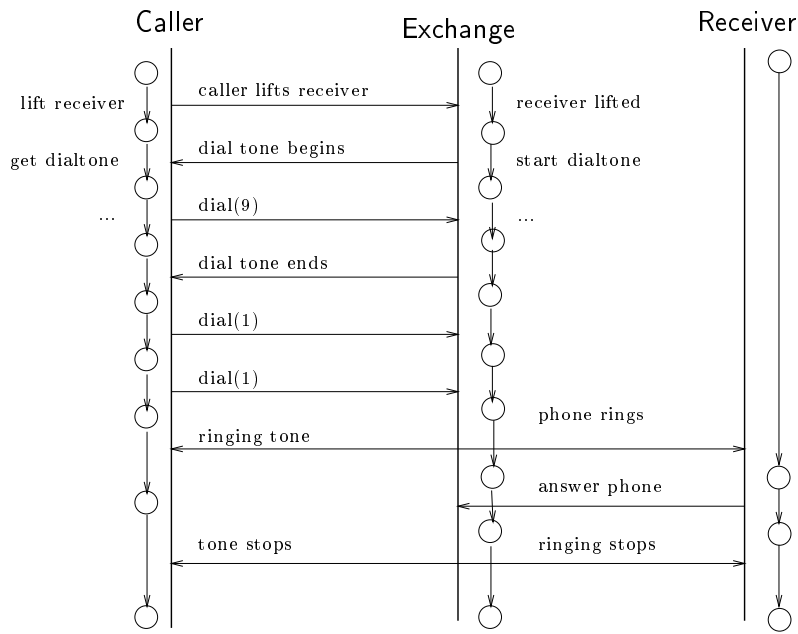

**Fig. 5.** Sequence diagram with abstract state automata

In contrast to the concrete state transition systems that are given by the
state diagrams of the involved objects and that describe the complete behavior
of the objects, the abstract state automata, which are derived from the sequence
diagram, only describe part of the behavior of the objects. These state automata
can be derived from the sequence diagram as follows:

– Between any two interactions, and before the first and after the last interaction, a state is introduced. Each abstract state $s_i^a$ of the state automaton corresponds to the set of concrete states $S_i^c$, which is a subset of the state space $states_{id}$ of the object. Note that the state sets corresponding to different abstract states do *not* have to be disjoint.
– With each interaction of the object in the sequence diagram, denoted by the $i$th arrow ending or beginning at the vertical line, an abstract transition between the states $s_i^a$ and $s_{i+1}^a$ is associated. This abstract transition corresponds to a nonempty set of concrete transitions of $\delta_{id}$, i.e. of the transition relation of the state transition system of the object (see Section 3.5).

The idea of using states between interactions is taken from [28]. In [28] "extended event traces" (EETs) are formalized. EETs are a notation similar to sequence diagrams; they are used with the objective to give a complete behavior description. Moreover, [8] shows how EETs can be used for describing complete interaction behavior in software architectures.

By using a state box view, our formalization makes it more apparent what is missing in sequence diagrams in order to be a specification technique:

– They leave completely open the relationship between abstract states in the sequence diagram and concrete states in the state diagrams of the involved objects.
– They only describe which concrete transitions may occur, but they do not forbid other concrete transitions.

To sum up, a sequence diagram describes the behavior of an object only partially, because it corresponds only to a subset of all paths in the state diagram of the object, and because it does not make this correspondence explicit. In contrast, a state diagram describes all paths, and, therefore, the complete behavior of the object.

### 3.5 State Diagrams

State diagrams serve as the connection between the structure of an object-oriented system and its behavior. Thus, state diagrams play a central role in the development of object-oriented systems. UML state diagrams look similar to Harel's statecharts [14]. However, several modifications and extensions make it difficult to define a precise semantics. In the following we sketch a semantic foundation based on the system model. The formalization is based on a semantic definition of similar state diagrams, which can be found in [12].

A state diagram can be attached either to a class or to the implementation of an operation ([2], Notation Guide, p.89). Their semantics differs accordingly. First we treat the semantics of class state diagrams. Class state diagrams are associated with the class names in *CN* and describe the lifecycles as well as the behavior of objects. The description is based on the actual state, which changes during the lifecycle.

**Class State Diagrams** In the following, we discuss the semantics of a state diagram associated with a class $c \in \mathit{CN}$ by transforming it into a state transition system (see Section 3.2).

**States** A class state diagram $c$ consists of a finite set $\mathit{STDStates}_c$ of possibly nested *diagram states* and a finite set $\mathit{STDTrans}_c$ of *diagram transitions*. Diagram states are (optionally) labeled by names that are taken from the set $\mathit{STDNames}_c$. A diagram state denotes an equivalence class of object states $\mathit{states}_{id}$ of the corresponding object. The semantics of elementary diagram states is, therefore, given by a function $st$ associating with each diagram state $S \in \mathit{STDStates}_c$ and each object identifier $id \in \mathit{ID}_c$ a corresponding set of object states $st(S, id) \subset \mathit{states}_{id}$.

The above requirement that each diagram state denotes exactly one equivalence class of object states can easily be achieved by assuming the name of the diagram state as an additional attribute of class $c$ (and introducing internal names for anonymous states).

The semantics of compound diagram states is defined as follows:

- The semantics of a composite diagram state (a so-called "OR-state") is given by the union of the state sets denoted by the subdiagram states.
- The semantics of a concurrently nested diagram state (a so-called "AND-state") is given by building the Cartesian product of its component diagram states.

Note that, although AND-states give a notion of concurrency, they can also be used to give a modular description of independent behavioral units of one sequential object. We do not allow feedback-composition of statecharts in order to simplify the semantic definition of state diagrams, as well as their understandability by the UML user. A similar comment was made in ([3], Metamodel, p.12).

As described above, the states of a state diagram are mapped to the states $\mathit{states}_{id}$ of the state transition system, which is already given by the semantics of a class diagram. The subset of initial states $\mathit{states}_c^0(id)$ is given by the states reachable by the initial event.

**Events** "An *event* is a significant occurrence. It has a location in time and space ..." ([2], Glossary, p.7). Therefore, we model events as simple transmissions of messages, occurring at some point in time. Each event $ev$ gives rise to a set of messages $msg(ev)$. Input events of class $c$ are modeled by the set $msg(ev) \subseteq msg_c$ of accepted messages. Similarly, output events are given as a subset of $\mathit{MSG}$.

UML distinguishes four different cases of events: receipt of a signal, receipt of an operation call, satisfaction of a condition and passage of a period of time ([2], Notation Guide, p.94). The first three are modeled as transitions, which are described in the next section. The semantics of the last is explained in [7] and not treated here.

**Transitions** "A ... *transition* is a relationship between two states ... when a specified event occurs ..." ([2], Notation Guide, p.96).

Each transition $(s, d, ev, out, C) \in STDTrans_c$ in the state diagram consists of a *source diagram state s*, a *destination diagram state d*, an input *event-signature ev*, a possibly empty output *send-clause out* and a guard condition $C$. UML also allows action-expressions, which are "... written in terms of operations, attributes, and links of the owning object ..." and "... must be an atomic operation." ([2], Notation Guide, p. 96). If the action expression does not contain calls or signals to other objects, it just restricts the resulting object states (and is, in this respect, similar to postconditions as allowed in Syntropy [10]). This can be easily incorporated into the semantics given below. However, when other objects are involved within an atomic action expression, communication with other objects is hidden in the action expression. As discussed below, in Section 3.5 on operation state diagrams, in a concurrent setting the semantics of communications not shown in the class state diagrams is not clear.

A transition in the diagram $(s, d, ev, out, C)$ is mapped to a set of transitions in the state transition system of the system model. Each transition in this set fulfills the following conditions:

- The transition starts in some state of the equivalence class $st_c(s)$ of the source diagram state and ends in some state of the equivalence class $st_c(d)$ of the destination diagram state.
- It is labeled with an input message from the set $msg(ev)$. This set may be empty.
- In addition, it is labeled with the set of output messages $msg(out)$. This set may be empty.
- It fulfills condition $C$.

The transition relation $\delta_c(id)$ of a state transition system of an object $id$ of class $c$ contains all transitions of these sets for all transitions in the state diagram.

UML distinguishes between simple transitions, complex transitions and transitions to nested states. We do not consider these details here, since composite states can always be expanded to simple states. We assume that the semantics of transitions is determined only after this expansion.

In addition to transitions, behavior can also be specified in UML state diagrams as internal activity, in particular entry, exit, and do actions. The latter can be treated similarly to general action expressions.

"If an event does not trigger any transitions, it is simply ignored." ([2], Notation Guide, p. 96). This is modeled by an extension of $\delta_c(id)$ with default transitions that leave the state unchanged. We remark, however, that another possibility is to model such events as chaotic behavior in the sense of *under-specification*. This allows for a refinement calculus on state diagrams as given in [26].

**Operation State Diagrams** It is difficult to define the semantics of a state diagram "... attached to a method (operation implementation) ..." ([2], Notation Guide, pp. 89), since none of the examples and only very little text in the UML documentation are devoted to this use. There are two major possiblities of how to associate a notion of state to an operation: either only the states of one object are shown such that the operation state diagram only describes the effect of the operation on one object, or the state covers several objects. In the latter case the diagram states must refer to a combination of the participating objects' states, thus modeling the "... condition during ... an interaction" ([2], Notation Guide, pp.90) this method is involved in. Furthermore, interactions between the participating objects are internal activities with respect to this operation state diagram.

In both cases the question arises, how several operation state diagrams and class diagrams should be combined and integrated. In a concurrent setting operation execution may be intertwined, such that not all states of each operation are visible in the object behavior. For example, the effect of a transfer operation between two bank accounts might not be visible in the object state after execution of the transfer operation, since concurrent deposits and withdrawals might have changed the accounts already.

Therefore, the simplest solution of combining class and operation state diagrams, namely, to view the operation state diagram just as as a complex action expression attached to the operation calls in the class state diagrams, is not always adequate. As an action expression, execution of operation state diagrams must be atomic (non-interruptible), which is not true for the transfer example above. In [4] a solution is discussed that attaches virtual objects to operation state diagrams, which can be called concurrently. This requires explicit synchronization of the access of the virtual operation objects to the object state. In [23] a solution is discussed that determines the semantics as the interleaving of the operation state diagrams based on a stack handling the operation calls. A thorough discussion of the different solutions is outside the scope of the paper. We just conclude that the combination of object behavior descriptions and operation behavior descriptions is an unsolved problem in the area of object-oriented modeling methods.

**General Remarks on State Diagrams** In the following, we suggest some improvements for state diagrams:

- In addition to *guard conditions*, postconditions should also be allowed. As mentioned above, this is a more abstract way of expressing the local effect of action expressions.
- There are several object-oriented approaches that implicitly use pattern matching, as used in functional programming languages, to relate input events and their argument values to the event triggers and their expressions. The use of these pattern matching techniques should be stated explicitly as a description mechanism in UML and be defined more precisely.

### 3.6 Activity Diagrams

Activity diagrams are a special case of state diagrams where all states have an internal action and no transition has an input event. They can be "... attached ... to a class or to the implementation of an operation and to a use case" ([2], Notation Guide, p.106). The first two cases have already been discussed for state diagrams in general (see section 3.5). In this section we discuss activity diagrams with swimlanes and action-object flow. These features seem to be particularly relevant for use case description. Another possible use would be to specify some operation of a composed object.

In the presence of swimlanes, the semantics of activity diagrams needs to be changed considerably. The main reason is that now several objects are involved and operate on their own object state. Thus, there is no notion of global state within one activity diagram and the transitions explicitly depict data and object flow between single activities. Hence, it is not adequate to give activity diagrams a semantics in terms of one state transition system.

As mentioned in ([2], Notation Guide, p.111), in some cases activity diagrams with action-object flow should be substituted by sequence diagrams. Also in our view, activity diagrams with swimlanes are more similar to sequence diagrams than to state diagrams. However, it is not clear from the UML documentation, whether they should only be used as a notational variant of sequence diagrams (where, for instance, action states correspond to named parts of the object lifeline) or whether some semantic differences are intended. Since they have not been included in earlier versions of UML [3], it seems likely that a more detailed explanation will be given in the next version.

## 4 Conclusion

In the preceding sections we have presented a proposal for the formal foundation of the Unified Modeling Language. As a direct result of our work, we detected a number of concepts that are not precisely defined, like the meaning of constraints in a concurrent setting of objects or the way how operations are specified and integrated in the overall object behavior. We also suggested enhancements of the UML descriptions, and we have argued that it is possible to map the UML language constructs to a coherent and sound semantic model.

A main idea of the semantics is to represent an overall system view in the semantic domain. This overall system view has been called system model. A system model describes both static and dynamic behavior of objects, including, for instance, dynamic object creation, concurrent behavior of objects with asynchronous message sending and inheritance relations.

The semantic domain of streams, on which our approach is based, has proved to be powerful enough to model specific properties of application domains like real-time systems and information systems. This is important, since UML claims to be an application independent analysis and design language.

There is still a lot of work to be done. Besides the precise elaboration of the semantics, there are several directions for future work.

A first main direction focuses on the benefits of the system model. As stated in the introduction, a formal semantics is the prerequisite for studying refinement steps, relationships between different description techniques, and for giving conditions that ensure the consistency of a system specification. In a second step, such properties have to be studied in the semantic domain, and, what is crucial, have to be formulated at the syntactical level of UML. Only if, for instance, consistency conditions can be formulated at the level of the description techniques, they can be integrated into a tool and support a sound system development. First work in this area has been presented, for instance in [26], where refinement steps for state diagrams are elaborated.

A second main direction for future work concerns aspects of the design process. Like UML itself, our semantic framework has been defined independently of a design methodology. Issues that still have to be addressed in more detail are, for instance, operation specifications and use case specifications. In the current stage of development, it is not clear what techniques effectively support the designer to specify operations and use cases and how they are integrated in the system specification. A first approach clarifying the relationships between the notions of messages, events and methods (operations) has been presented in [4]. These studies provide guidelines and schemes for integrating partial views of a system (like operation behavior) into an overall system view and assist the developer to gain a structured and sound system specification.

**Acknowledgments**

We thank Grady Booch, Ivar Jacobson and Gunnar Övergaard for interesting discussions regarding UML. We also thank Manfred Broy and Ingolf Krüger for stimulating discussions and comments on earlier versions of this paper.